\documentclass[a4paper,fleqn]{cas-dc}

\usepackage[numbers]{natbib}
\usepackage{multirow}
\usepackage{subfig}
\usepackage{wrapfig}
\usepackage{balance}
\def\tsc#1{\csdef{#1}{\textsc{\lowercase{#1}}\xspace}}
\tsc{WGM}
\tsc{QE}

\begin{document}
\let\WriteBookmarks\relax
\def\floatpagepagefraction{1}
\def\textpagefraction{.001}
\let\printorcid\relax

\shorttitle{GraphMFT: A graph network based multimodal fusion technique}

\shortauthors{Li et al.}  

\title [mode = title]{GraphMFT: A graph network based multimodal fusion technique for emotion recognition in conversation}

\tnotetext[1]{Accepted by Neurocomputing. DOI: 10.1016/j.neucom.2023.12642. Received 9 August 2022; Revised 3 April 2023; Accepted 4 June 2023. \textit{E-mail address:} lijfrank@hust.edu.cn (J. Li), wangxiaoping@ hust.edu.cn (X. Wang).}

\author[1,2,3]{Jiang Li}
\author[1,2,3]{Xiaoping Wang}
\cormark[1]
\cortext[1]{Corresponding author at: School of Artificial Intelligence and Automation, Huazhong University of Science and Technology, Wuhan 430074, China.}
\author[1,2,3]{Guoqing Lv}
\author[1,2,3]{Zhigang Zeng}
\address[1]{School of Artificial Intelligence and Automation, Huazhong University of Science and Technology (HUST), Wuhan 430074, China}
\address[2]{Key Laboratory of Image Processing and Intelligent Control, Ministry of Education, Wuhan 430074, China}
\address[3]{Hubei Key Laboratory of Brain-inspired Intelligent Systems, Wuhan 430074, China}

\begin{abstract}
Multimodal machine learning is an emerging area of research, which has received a great deal of scholarly attention in recent years. Up to now, there are few studies on multimodal Emotion Recognition in Conversation (ERC). Since Graph Neural Networks (GNNs) possess the powerful capacity of relational modeling, they have an inherent advantage in the field of multimodal learning. GNNs leverage the graph constructed from multimodal data to perform intra- and inter-modal information interaction, which effectively facilitates the integration and complementation of multimodal data. In this work, we propose a novel Graph network based Multimodal Fusion Technique (GraphMFT) for emotion recognition in conversation. Multimodal data can be modeled as a graph, where each data object is regarded as a node, and both intra- and inter-modal dependencies existing between data objects can be regarded as edges. GraphMFT utilizes multiple improved graph attention networks to capture intra-modal contextual information and inter-modal complementary information. In addition, the proposed GraphMFT attempts to address the challenges of existing graph-based multimodal conversational emotion recognition models such as MMGCN. Empirical results on two public multimodal datasets reveal that our model outperforms the State-Of-The-Art (SOTA) approaches with the accuracy of 67.90\% and 61.30\%.
\end{abstract}
\maketitle

\section{Introduction}\label{introduction}
As is known to all, the prerequisite for an empathic response is that we can accurately recognize the other party's emotions. Recently, the applications of Human-Computer Interaction (HCI) have received widespread attention. The ability that machine can provide services naturally like a human is a pressing issue. Inspired by the empathy that human generates, machines should be endowed with the ability to detect emotions before providing services. With the synergistic development of artificial intelligence and cognitive science, Emotion Recognition in Conversation (ERC) is playing an outstanding role in numerous applications of HCI. Examples include conversational robot~\cite{sabelli2011conversational}, opinion mining~\cite{cortis2021over}, medical diagnosis~\cite{bhavan2019bagged}.

Most ERC approaches model interactions based on intra-modal context. HiGRU~\cite{jiao2019higru} leverages two improved Gated Recurrent Units (GRUs) to model word-level information and utterance-level context, respectively, and achieves outstanding performance on three dialogue emotion datasets. DialogueRNN~\cite{majumder2019dialoguernn} updates contextual information of the utterance and state of the participant through global GRU and party GRU, respectively, while emotion GRU is used for decoding emotion representation of the utterance. COSMIC~\cite{ghosal2020cosmic} improves DialogueRNN and introduces external knowledge. Hu et al.~\cite{hu2021dialoguecrn} are motivated by Cognitive Theory of Emotion~\cite{lazarus1991progress} to model contextual information flow from a cognitive reasoning perspective. However, these recurrence-based methods have difficulty capturing information of long-distance utterances and focus on considering the nearest ones to update state of query utterance. Since Kipf et al.~\cite{kipf2016semi} proposed Graph Convolutional Network (GCN), Graph Neural Networks (GNNs) have been frequently applied to computer vision, natural language processing, recommendation systems, etc. GNNs can easily capture long-distance contextual information in ERC tasks through the powerful capability of relational modeling. DialogueGCN~\cite{ghosal2019dialoguegcn} models each conversation as a graph, which addresses the problem of context propagation in the recurrence-based methods. Ishiwatari et al.~\cite{ishiwatari2020relation} consider both the speaker dependency and the sequential information by combining graph neural networks and relational position encodings. KET~\cite{zhong2019knowledge} adopts a dynamic context-aware graph attention mechanism to capture external commonsense knowledge and promote the understanding of each word in the utterances. DAG-ERC~\cite{DBLP:conf/acl/ShenWYQ20} models long-distance contextual information flow through a directed acyclic graph for emotion detection. Unfortunately, the recurrence-based and graph-based approaches mentioned above either neglect multimodal information or fail to consider inter-modal interaction information.

\begin{figure}[htbp]
    \centering
    \includegraphics[width=3.0in]{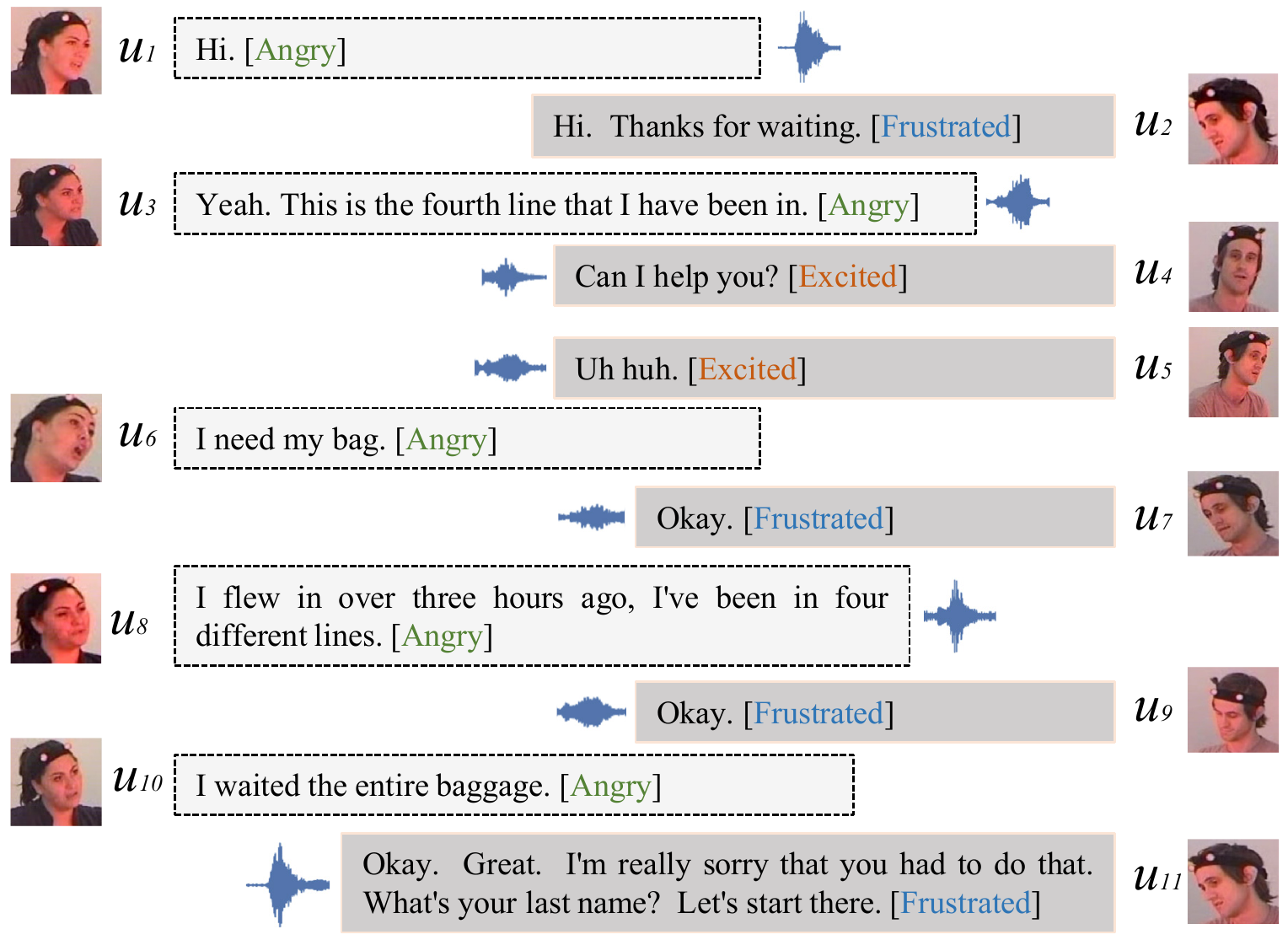}
    \caption{A multimodal conversation scenario consisting of visual, textual and acoustic modalities.}
    \label{fig:example}
\end{figure}
Multimodal machine learning has risen in popularity in recent years. Multimodality is originally proposed to compensate for the deficiency that unimodality does not sufficiently reflect the reality state in some circumstances. Multimodal ERC is influenced by the fact that humans perceive the world from multisensory information. Figure~\ref{fig:example} illustrates a multimodal dialogue scenario whose sources mainly include the visual, acoustic and textual modalities. ConGCN~\cite{zhang2019modeling} models both utterances and speakers as nodes, as well as both context dependencies and speaker dependencies as edges, for multi-participant conversational emotion detection. But ConGCN only takes into account textual features and acoustic features. MMGCN~\cite{hu2021mmgcn} considers both multimodal information and long-distance contextual information through graph-based model, and shows its superiority over the baseline models on two multimodal datasets. Nevertheless, MMGCN coarsely connects the utterance to all other intra-modal utterances, which brings additional noise. Although GNNs have a high capability of modeling relations, which demands that the input to the model is a homogeneous graph~\cite{wu2020comprehensive}. MMGCN integrates the utterance nodes of three modalities into a single GNN simultaneously, which coarsely treats the complex heterogeneous graph as a homogeneous graph as the direct input of GNN. Furthermore, Multimodal fusion bears the problem of data heterogeneity~\cite{jia2021hetemotionnet}, and the more types of modality there are, the more difficult it is to fuse them.

In order to tackle the mentioned dilemmas, we present a novel multimodal fusion method based on graph neural networks, namely Graph network based Multimodal Fusion Technique (GraphMFT), for conversational emotion detection in this paper. GraphMFT utilizes multiple improved Graph ATtention networks (GATs) to extract intra-modal contextual dependencies and inter-modal complementary dependencies, which can effectively facilitate the interaction of current utterance with intra-modal and inter-modal utterances. The proposed GraphMFT differs significantly from MMGCN, mainly including: (1) We construct three graphs, each containing two modalities. GraphMFT employs three improved GATs, each only fusing information from two modalities. So we mitigate the problem of data heterogeneity in multimodal fusion techniques by reducing the number of modalities in the network. (2) Instead of connecting the current node to all nodes within the modality, we connect it to variable context nodes, cancelling out the potential for additional noise. (3) We adopt a spatial-based GNN as the basic backbone network, which requires explicit construction of edges, while MMGCN adopts a spectral-based GNN, which is not necessary to construct edges explicitly, but only a big adjacency matrix. (4) GraphMFT leverages the graph attention module to learn the weights of edges, while MMGCN utilizes the angular similarity to represent these weights. (5) We design a new fusion network, while MMGCN directly applies the off-the-shelf network structure, i.e., GCNII~\cite{chen2020simple}.

Our main contributions in this work can be summarized as follows:
\begin{enumerate}[$(1)$]
    \item A graph-based multimodal fusion model, GraphMFT, is proposed. GraphMFT can effectively capture intra-modal contextual information and inter-modal complementary information. Meanwhile, since there are few multimodal ERC models based on GNNs, we contribute a new fusion architecture for multimodal machine learning.
    \item We model a conversation as three graphs, i.e., $\mathrm{V}$-$\mathrm{A}$ graph, $\mathrm{V}$-$\mathrm{T}$ graph, and $\mathrm{A}$-$\mathrm{T}$ graph. Each graph contains information from only two modalities, which decreases the difficulty of multimodal fusion. The nodes of each graph are the utterances of each modality in the conversation, and the edges of each graph include both intra-modal dependencies and inter-modal dependencies.
    \item We conduct extensive experiments on two public benchmark datasets, i.e., IEMOCAP and MELD datasets. Experimental results show that our model outperforms all baseline methods.
\end{enumerate}

The remaining of this paper is organized as follows. In Section~\ref{sec:architecture}, the architecture we propose is presented with a focus on graph network based multimodal fusion techniques. In Section~\ref{sec:experiments}, the experimental preparation and experimental setup are described. We next report the experimental results with an in-depth discussion and analysis in Section~\ref{sec:results}. Besides, Section~\ref{sec:related} and Section~\ref{sec:summary} are an introduction to related works and a summary of this paper, respectively.

\section{Related works}\label{sec:related}
\subsection{Emotion recognition in conversion}
Emotion Recognition in Conversation (ERC) aims to determine the emotion state of each utterance in a conversation and plays a crucial role in affective dialogue systems. How to model the conversational context and multimodal interaction is the core of this task. According to the number of input modality, existing ERC models can be divided into unimodal approaches and multimodal approaches.

\textbf{Unimodal Approaches} utilize single modal input information to judge the emotion of each utterance. Current unimodal-based approaches focus on utilizing textual modality for conversational emotion detection. KET~\cite{zhong2019knowledge} exploits hierarchical self-attention to interpret contextual utterances and utilizes a context-aware graph attention mechanism to dynamically learn external commonsense knowledge, which improves the performance of ERC task. DialogXL~\cite{shen2021dialogxl} addresses the issue that the pre-trained language models can't be applied in the hierarchical structures of conversational utterances. COSMIC~\cite{ghosal2020cosmic} utilizes commonsense including mental states, events, and causal relations to learn interactions between interlocutors, which addresses challenges such as emotion shift detection and context propagation. In order to model the intrinsic structure of a conversation, DAG-ERC~\cite{DBLP:conf/acl/ShenWYQ20} encodes the utterances with a directed acyclic graph in an attempt to combine the advantages of graph neural networks and recurrent neural networks. Other unimodal ERC methods include HiGRU~\cite{jiao2019higru}, DialogueGCN~\cite{ghosal2019dialoguegcn}, and DialogueCRN~\cite{hu2021dialoguecrn}. With the advancement of multimodal machine learning and the emergence of increasingly complex conversational sentiment recognition scenarios, the unimodal approaches offer limited performance gains.

\textbf{Multimodal Approaches} detect emotions in conversations through multimodal information. As the information of unimodality is biased and susceptible to interference from external noise factors, the research on ERC models based on multimodal fusion based has been receiving widespread attention for their merit of the complementarity between different modalities~\cite{poria2017review}. CMN~\cite{hazarika2018conversational} utilizes multimodal information by means of directly concatenating three modalities features, and adopts the Gated Recurrent Unit (GRU) to model contextual information. ICON~\cite{hazarika2018icon} extracts multimodal conversation features and leverages global memories to model emotional influences of the self- and inter-speakers hierarchically, which generates contextual summaries to improve the performance of utterance-videos emotion recognition. ConGCN~\cite{zhang2019modeling} is a GCN-based multimodal fusion model for ERC task. In this method, the node represents an utterance or a speaker, the edge between two utterances represents the context-sensitive dependence, and the edge between a speaker and its corresponding utterance represents the speaker-sensitive dependence. MMGCN~\cite{hu2021mmgcn} is also a GCN-based model for the multimodal ERC task, which can capture not only long-distance contextual information but also multimodal interactive information. Other multimodal ERC methods include MFN~\cite{zadeh2018memory}, BC-LSTM~\cite{poria2017context}, and DialogueRNN~\cite{majumder2019dialoguernn}. These methods either do not capture the interaction information between modalities or ignore the heterogeneity of multimodal data. Additionally, there are few multimodal-based ERC models, and more relevant research is necessary to advance the field of ERC. Based on this, we propose a new GNN-based multimodal fusion method for conversational emotion detection. The proposed model is intended to minimize the problems mentioned above, as well as to facilitate the development of the multimodal learning field.

\subsection{Graph convolutional networks}
Graphs naturally exist in many practical applications, such as social analysis and fraud detection. The data is represented as graphs that can encode the structural information to model the relations among entities. However, the irregular structure of the graph data makes it difficult to define convolution and filtering on graphs. To this end, Graph Convolutional Networks (GCNs) are put forward to conduct convolutional operations on the graph. GCNs have been playing an essential role in numerous fields, such as natural language processing, computer vision, recommender system. GCNs can be mainly categorized into spectral models and spatial models. Spectral models are defined based on graph Fourier Transform, while spatial models are defined as the aggregations of neighboring nodes. The widely-used GCNs include 1stChebNet~\cite{kipf2016semi}, GraphSAGE~\cite{hamilton2017inductive}, GAT~\cite{velickovic2017graph}. In this work, unlike MMGCN which employs spectral GCN, we leverage a spatial-based approach (i.e. GAT) as the base backbone network.

\subsection{Multimodal machine learning}
The goal of multimodal machine learning is to relate and process information from multiple modalities, which is extraordinarily potential in many fields and attaching increasingly more importance. It is composed of primarily five parts: representation, translation, alignment, fusion, and co-learning, where multimodal fusion is one of the most researched topics~\cite{baltruvsaitis2018multimodal}. CM-BERT~\cite{yang2020cm} proposes a cross-modal BERT approach to fine-tune the pre-trained BERT model through the interaction of text and audio modalities. Ephrat et al.~\cite{ephrat2018looking} propose a deep network-based framework which combines visual and auditory signals to separate a single speech signal from mixture sounds such as ambient noise or other speakers. Multilogue-Net~\cite{shenoy2020multilogue} presents an end to end Recurrent Neural Network (RNN) architecture to extract the contextual information of all modalities in the dialogue and capture interactions between the available modalities. Sahay et al.~\cite{sahay2018multimodal} adopt the inter-modal interactions of a segment while taking into account the sequence of segments in a video to model the inter-segment inter-modal interactive relationship. Kumar et al.~\cite{kumar2020gated} suggest that learning cross-modal interactions is instrumental in solving the problems confronting multimodal sentiment analysis, and propose a gating mechanism for multimodal interaction and fusion. GME-LSTM~\cite{chen2017multimodal} fuses multimodal information at the word level and leverages the gated multimodal embedding to alleviate the difficulties of fusing noisy modalities, which can effectively model the multimodal sequences of speech and perform sentiment comprehension. BBFN~\cite{han2021bi} fuses relevance increment and separate difference increment of pairwise modalities, and trains simultaneously the two parts to simulate the combat between them. Graph neural networks possess tremendous ability to model relationships and extract interactive information between different modalities. However, to our knowledge, there are still relatively few existing GNN-based multimodal fusion models. Therefore, we aspire to contribute a new GNN-based architecture to advance the development of multimodal learning.

\section{The proposed architecture}\label{sec:architecture}
\begin{figure*}[htbp]
    \centering
    \includegraphics[width=6.0in]{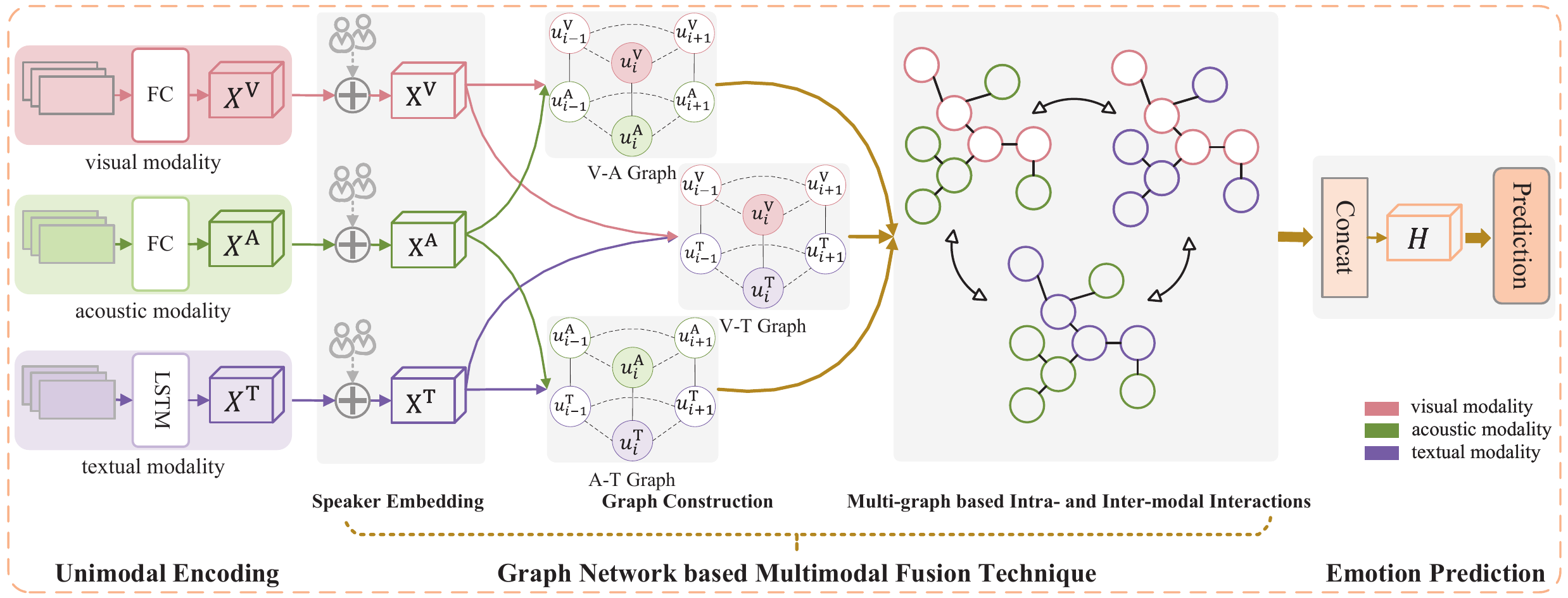}
    \caption{The overall architecture of our approach, which consists of unimodal encoding, graph network based multimodal fusion technique, and emotion prediction. Here, the flow diagram of multi-graph based intra- and inter-modal interactions can be seen in Figure~\ref{fig:interactions}.}
    \label{fig:overall}
\end{figure*}
This section contains a detailed presentation of method proposed in this paper. The overall architecture of our approach is shown in Figure~\ref{fig:overall}, including unimodal encoding, graph network based multimodal fusion technique, and emotion prediction. Here, graph network based multimodal fusion technique mainly consists of speaker embedding, graph construction, and multi-graph based intra- and inter-modal interactions.

\subsection{Problem definition}
In the multimodal ERC system, each conversation contains $m$ utterances $u_1$, $u_2$, $\cdots$, $u_m$, and each utterance $u_i$ has 3 modal expressions $u_i^\mathrm{V}$, $u_i^\mathrm{A}$, $u_i^\mathrm{T}$. Thus, there exist a total of $3 \times m$ utterance expressions. Here, $\mathrm{V}$, $\mathrm{A}$, and $\mathrm{T}$ denote visual, acoustic, and textual modalities, respectively; we refer to $u_i^\mathrm{V}$, $u_i^\mathrm{A}$, and $u_i^\mathrm{T}$ as visual, acoustic, and textual utterances, respectively. Each utterance $u_i$ (containing three modalities, i.e., $u_i^\mathrm{V}$, $u_i^\mathrm{A}$, $u_i^\mathrm{T}$) corresponds to an emotion $y_i$. Each conversation involves two and more speakers, so two utterances may be from the same speaker or from different speakers. Given utterance $u_i$, the goal of multimodal ERC task is to predict emotion $y_i$ corresponding to $u_i$ based on intra- and inter-modal interaction learning. The multimodal emotion datasets used include IEMOCAP, which includes 6 emotion labels, and MELD, which includes 7 emotion labels. Referring to the approach of MMGCN~\cite{hu2021mmgcn}, the features of visual, acoustic and text modalities are extracted employing DenseNet~\cite{huang2017densely}, OpenSmile toolkit~\cite{schuller2011recognising} and TextCNN~\cite{kim-2014-convolutional}, respectively.

\subsection{Unimodal encoding}
We exploit fully connected networks to encode utterances in the visual and acoustic modalities separately to enhance the expressiveness of utterance features. For textual modal utterances, a bidirectional Long Short-Term Memory (LSTM) network is leveraged to extract the contextual information of current utterance. The process of unimodal encoding can be described by the following equations:
\begin{equation}
    \label{eq:pre_encoding}
    \begin{split}
    &x_i^\mathrm{V} = \mathrm{W}_e^\mathrm{V} c_i^\mathrm{V} + \mathrm{b}_{e}^\mathrm{V},\\
    &x_i^\mathrm{A} = \mathrm{W}_e^\mathrm{A} c_i^\mathrm{A} + \mathrm{b}_{e}^\mathrm{A},\\
    &x_i^\mathrm{T}, x_{h,i}^\mathrm{T} = \overleftrightarrow{\mathrm{LSTM}}_e(c_i^\mathrm{T}, x_{h,i-1}^\mathrm{T}, x_{h,i+1}^\mathrm{T}),
    \end{split}
\end{equation}
where $c_i^\mathrm{V}$, $c_i^\mathrm{A}$, and $c_i^\mathrm{T}$ are inputs of unimodal encoding; $x_i^\mathrm{V}$, $x_i^\mathrm{A}$, $x_i^\mathrm{T}$ are outputs of unimodal encoding; $\mathrm{V}$, $\mathrm{A}$, and $\mathrm{T}$ denote visual, acoustic, and textual modalities, respectively; $\overleftrightarrow{\mathrm{LSTM}}_e$ denotes the bidirectional LSTM network; $\mathrm{W}_e^\mathrm{V}$, $\mathrm{W}_e^\mathrm{A}$, $\mathrm{b}_{e}^\mathrm{V}$, and $\mathrm{b}_{e}^\mathrm{A}$ are the trainable parameters.

\subsection{Graph network based multimodal fusion technique}
In order to encode both intra-modal contextual information and inter-modal complementary information, we propose a Graph network based Multimodal Fusion Technique (GraphMFT) for Emotion Detection in Conversation. The proposed GraphMFT allows an utterance of current modality to interact with both intra-modal utterances and inter-modal utterances at the same time.

\subsubsection{Speaker embedding}
Referring to Hu et al.~\cite{hu2021mmgcn}, we encode speakers as embedding vectors to capture the characteristics of speakers. The speaker embedding is then summed with utterance feature of each modality to obtain new utterance feature carrying speaker information. The above process can be represented as follows:
\begin{equation}
    \label{eq:emb-spk}
    \begin{split}
    &S_{pk} = \mathrm{Embedding}(\mathrm{SPK}, s),\\
    &\mathrm{X}^\tau = X^\tau + S_{pk}, \tau \in \{\mathrm{V},\mathrm{A},\mathrm{T}\},
    \end{split}
\end{equation}
where $s$ is the number of speakers, and $\mathrm{SPK}$ is the speaker array (note that different speakers are represented as different index values); $\mathrm{Embedding}$ denotes the embedding layer, and $S_{pk}$ is the speaker embedding obtained by feeding the speaker array and the number of speakers to the embedding layer; $X^\tau$ is the feature set after unimodal encoding ($x_i^\tau \in X^\tau$), $\mathrm{X}^\tau$ denotes the feature set adding speaker information, and $\mathrm{V}$, $\mathrm{A}$, $\mathrm{T}$ indicate visual, acoustic, textual modalities, respectively.

\subsubsection{Graph construction}
We construct three graphs for each dialogue, including $\mathrm{V}$-$\mathrm{A}$ graph, $\mathrm{V}$-$\mathrm{T}$ graph, and $\mathrm{A}$-$\mathrm{T}$ graph. These graphs can be represented as $\mathcal{G}^\mathrm{VA} = (\mathcal{V}^\mathrm{VA}, \mathcal{E}^\mathrm{VA}, \mathcal{W}^\mathrm{VA})$, $\mathcal{G}^\mathrm{VT}=(\mathcal{V}^\mathrm{VT}, \mathcal{E}^\mathrm{VT}, \mathcal{W}^\mathrm{VT})$, and $\mathcal{G}^\mathrm{AT}=(\mathcal{V}^\mathrm{AT}, \mathcal{E}^\mathrm{AT}, \mathcal{W}^\mathrm{AT})$. Here, $\mathcal{G}^\mathrm{VA}$, $\mathcal{G}^\mathrm{VT}$, and $\mathcal{G}^\mathrm{AT}$ denote $\mathrm{V}$-$\mathrm{A}$ graph, $\mathrm{V}$-$\mathrm{T}$ graph, and $\mathrm{A}$-$\mathrm{T}$ graph, respectively; $\mathcal{V}^\mathrm{VA}$, $\mathcal{V}^\mathrm{VT}$, and $\mathcal{V}^\mathrm{AT}$ denote the node sets, of which each element (i.e., node) denotes an utterance of a certain modality; for instance, $u_i^\mathrm{V}$ and $u_i^\mathrm{A}$ denote the $i$-th visual utterance and acoustic utterance, respectively; $\mathcal{E}^\mathrm{VA}$, $\mathcal{E}^\mathrm{VT}$, and $\mathcal{E}^\mathrm{AT}$ denote the edge sets between nodes; $\mathcal{W}^\mathrm{VA}$, $\mathcal{W}^\mathrm{VT}$, and $\mathcal{W}^\mathrm{AT}$ denote the edge weights. Specifically, our graphs is constructed by the following way.

$\mathbf{V}$-$\mathbf{A}$ \textbf{Graph}, as the name suggests, concerns visual and acoustic modalities of each utterance in a conversation. 
\\ \noindent 
\textit{\textbf{Nodes}}: In a conversation, each visual and acoustic utterance is treated as a node in $\mathrm{V}$-$\mathrm{A}$ graph. So there exist a total of $2 \times m$ nodes, where $m$ is the number of utterances of current modality in a conversation. 
\\ \noindent 
\textit{\textbf{Edges}}: We create two types of edges in multimodal scenarios, i.e., intra-modal edges and inter-modal edges. Each node is connected to the past $\mathcal{P}$ or future $\mathcal{F}$ contextual nodes of current modality, i.e., intra-modal edges are created. In addition, we connect each node with nodes belonging to different modalities but from the same utterance, i.e., inter-modal edges are created. For example, $u_i^\mathrm{V}$ is connected not only to the past $\mathcal{P}$ (future $\mathcal{F}$) nodes $u_{i-\mathcal{P}}^\mathrm{V}$, $\cdots$, $u_{i-1}^\mathrm{V}$ ($u_{i+1}^\mathrm{V}$, $\cdots$, $u_{i+\mathcal{F}}^\mathrm{V}$) but also to node $u_i^\mathrm{A}$. 
\\ \noindent 
\textit{\textbf{Edge Weights}}: Distinct neighboring nodes may impact current utterance node differently. To extract the importance of different neighboring nodes, we adopt graph attention module~\cite{velickovic2017graph} to calculate the edge weights. Specifically, the edge weight of $u_i$ and $u_j$ can be calculated as:
\begin{equation}
    \label{eq:gat}
    \omega_{ij} =
    \frac{
    \exp\left(\sigma\left({\mathrm{a}_\omega}^{\top}
    [\mathrm{W}_\omega \mathrm{x}_i \, \Vert \, \mathrm{W}_\omega \mathrm{x}_j]
    \right)\right)}
    {\sum_{u_k \in \mathcal{N}(u_i)}
    \exp\left(\sigma\left({\mathrm{a}_\omega}^{\top}
    [\mathrm{W}_\omega \mathrm{x}_i \, \Vert \, \mathrm{W}_\omega \mathrm{x}_k]
    \right)\right)},
\end{equation}
where $\mathrm{x}_i \in \mathrm{X}^\mathrm{V} \cup \mathrm{X}^\mathrm{A}$, and $\mathrm{x}_i$ is the feature representation of node $u_i$, and $u_i \in \mathcal{V}^\mathrm{VA}$; $\mathrm{x}_j$ and $\mathrm{x}_k$ is the feature representation of $u_j$ and $u_k$, respectively; Both $u_j$ and $u_k$ are neighboring nodes of $u_i$; $\omega_{ij}$ denotes the edge weight of $u_i$ and $u_j$, and $\sigma$ indicates non-linear activation function $\mathrm{LeakyReLU}$; $\mathrm{W}_\omega$ and $\mathrm{a}_\omega$ are the learnable parameters.

$\mathbf{V}$-$\mathbf{T}$ \textbf{Graph} and $\mathbf{A}$-$\mathbf{T}$ \textbf{Graph} are constructed in the same way as $\mathrm{V}$-$\mathrm{A}$ graph. Here, $\mathrm{V}$-$\mathrm{T}$ graph concerns visual modality and textual modality, and $\mathrm{A}$-$\mathrm{T}$ graph concerns acoustic modality and textual modality.

\subsubsection{Multi-graph based intra- and inter-modal interactions}\label{sec:multi-graph_interactions}
\begin{figure}[htbp]
    \centering
    \includegraphics[width=3.3in]{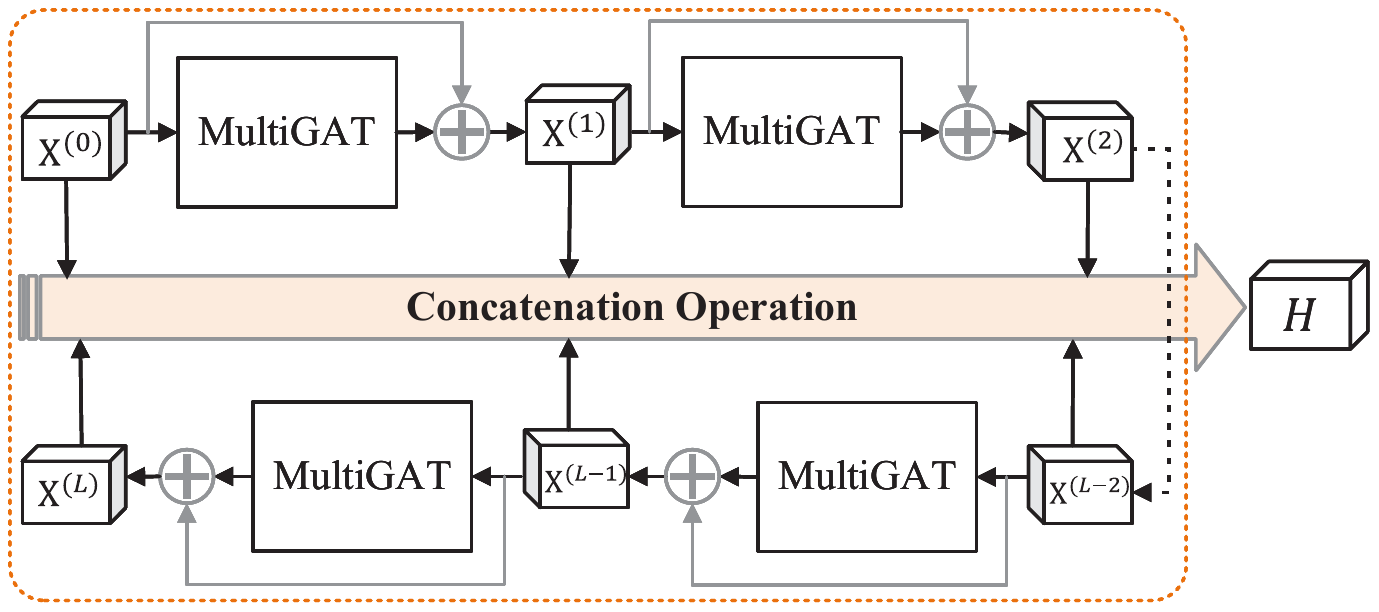}
    \caption{The network structure of improved GAT, where $\mathrm{MultiGAT}$ denotes multi-head graph attention layer.}
    \label{fig:improved}
\end{figure}
It has been proven that GNNs possess powerful capability to model relations. In this work, we model multimodal ERC with the help of Graph ATtention networks (GAT)~\cite{velickovic2017graph}. Inspired by ResGCN~\cite{li2021deepgcns}, we make modifications to the vanilla GAT to alleviate the over-smoothing problem of GNN. Figure~\ref{fig:improved} is an illustration of improved GAT. Specifically, we connect the previous layer network's output to the following layer's output. In other words, the output of the previous layer network is added to the output of the next layer. Not only that, we also keep the outputs of all layers and then concatenate them to obtain the final result. The improved GAT be formalized as:
\begin{equation}
    \label{eq:improved}
    \begin{split}
    &\mathrm{X}^{(l)} = \mathrm{X}^{(l-1)}+\mathrm{MultiGAT}(\mathrm{X}^{(l-1)},\mathcal{E}),\\
    &H = \mathrm{W}_{im}[\mathrm{X}^{(0)} \Vert \mathrm{X}^{(1)} \Vert\cdots\Vert \mathrm{X}^{(L)}],
    \end{split}
\end{equation}
where $l=1,\cdots,L$, and $L$ is the number of network layers; $\mathrm{MultiGAT}$ denotes the multi-head graph attention layer; $\mathrm{X}^{(l-1)}$ is the $(l-1)$-th layer feature matrix, and $\mathcal{E}$ is the edge set, and they are both inputs to GAT; $H$ is the output of GAT, $\mathrm{W}_{im}$ is the trainable parameter.

\begin{figure}[htbp]
    \centering
    \includegraphics[width=3.0in]{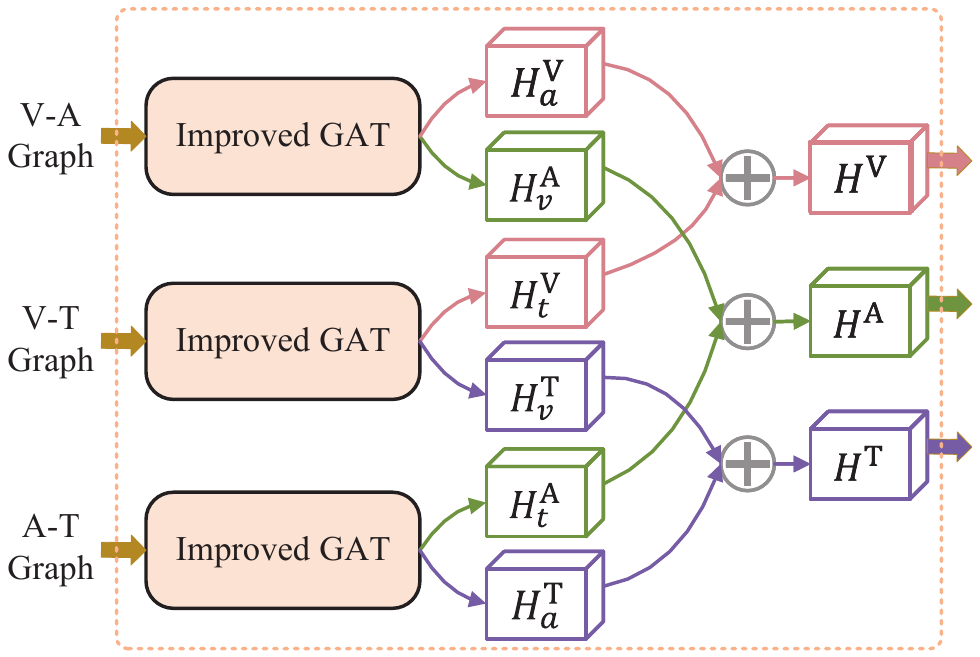}
    \caption{Illustration of multi-graph based intra- and inter-modal interactions. The network structure of improved GAT can be seen in Figure~\ref{fig:improved}.}
    \label{fig:interactions}
\end{figure}
We employ three GATs in order to extract both intra- and inter-modal interaction information, where the input of each GAT includes information from two modalities. As shown in Figure~\ref{fig:interactions}, we take $\mathrm{V}$-$\mathrm{A}$ graph, $\mathrm{V}$-$\mathrm{T}$ graph, and $\mathrm{A}$-$\mathrm{T}$ graph as the inputs of improved GATs; then the feature matrices of intra- and inter-modal information interactions are obtained. Taking node $u_i^\mathrm{V}$ in $\mathrm{V}$-$\mathrm{A}$ graph as an example, $u_i^\mathrm{V}$ collects both visual contextual information (i.e., the information of $u_{i-\mathcal{P}}^\mathrm{V}$, $\cdots$, $u_{i-1}^\mathrm{V}$, $u_{i+1}^\mathrm{V}$, $\cdots$, $u_{i+\mathcal{F}}^\mathrm{V}$) and acoustic complementary information (i.e., $u_i^\mathrm{A}$) through GATs. We can describe computational procedure of $\mathrm{V}$-$\mathrm{A}$ graph as:
\begin{equation}
    \label{eq:interaction0}
    \begin{split}
    &\mathrm{X}^{\mathrm{VA},(l)} = \mathrm{X}^{\mathrm{VA},(l-1)}+\mathrm{MultiGAT}(\mathrm{X}^{\mathrm{VA},(l-1)},\mathcal{E}^\mathrm{VA}),\\
    &H^\mathrm{VA} = \mathrm{W}_{va}[\mathrm{X}^{\mathrm{VA},(0)} \Vert \mathrm{X}^{\mathrm{VA},(1)} \Vert \cdots \Vert \mathrm{X}^{\mathrm{VA},(L)}],
    \end{split}
\end{equation}
where $l=1,\cdots,L$, and $L$ is the number of layers of improved GAT; $\mathrm{X}^{\mathrm{VA},(l)}=\mathrm{X}^{\mathrm{V},(l)} \cup \mathrm{X}^{\mathrm{A},(l)}$, and $\mathrm{X}^{\mathrm{VA},(l)}$ is the $l$th layer node feature matrix in $\mathrm{V}$-$\mathrm{A}$ graph; $\mathrm{X}^{\mathrm{VA},(0)}=\mathrm{X}^{\mathrm{V}} \cup \mathrm{X}^{\mathrm{A}}$ is the input of improved GAT, $H^\mathrm{VA}$ is the output of improved GAT; $\mathcal{E}^\mathrm{VA}$ is the edge set in $\mathrm{V}$-$\mathrm{A}$ graph, and $\mathrm{MultiGAT}$ is multi-head graph attention layer; $\Vert$ denotes the concatenation operation, and $\mathrm{W}_{va}$ is the learnable parameter.

Similarly, according to the computational procedure of $\mathrm{V}$-$\mathrm{A}$ graph, $\mathrm{V}$-$\mathrm{T}$ graph and $\mathrm{A}$-$\mathrm{T}$ graph are calculated as:
\begin{equation}
    \label{eq:interaction1}
    \begin{split}
    &\mathrm{X}^{\mathrm{VT},(l)} = \mathrm{X}^{\mathrm{VT},(l-1)}+\mathrm{MultiGAT}(\mathrm{X}^{\mathrm{VT},(l-1)},\mathcal{E}^\mathrm{VT}),\\
    &H^\mathrm{VT} = \mathrm{W}_{vt}[\mathrm{X}^{\mathrm{VT},(0)} \Vert \mathrm{X}^{\mathrm{VT},(1)} \Vert \cdots \Vert \mathrm{X}^{\mathrm{VT},(L)}],\\
    &\mathrm{X}^{\mathrm{AT},(l)} = \mathrm{X}^{\mathrm{AT},(l-1)}+\mathrm{MultiGAT}(\mathrm{X}^{\mathrm{AT},(l-1)},\mathcal{E}^\mathrm{AT}),\\
    &H^\mathrm{AT} = \mathrm{W}_{at}[\mathrm{X}^{\mathrm{AT},(0)} \Vert \mathrm{X}^{\mathrm{AT},(1)} \Vert \cdots \Vert \mathrm{X}^{\mathrm{AT},(L)}],
    \end{split}
\end{equation}
where $H^\mathrm{VT}$ and $H^\mathrm{AT}$ are the outputs of two improved GATs; $\mathrm{W}_{vt}$ and $\mathrm{W}_{at}$ are the learnable parameters. Then, we sum feature matrices of the same modalities,
\begin{equation}
    \label{eq:sum}
    \begin{split}
        &H^\mathrm{V} = H_a^\mathrm{V}+H_t^\mathrm{V},\\
        &H^\mathrm{A} = H_v^\mathrm{A}+H_t^\mathrm{A},\\
        &H^\mathrm{T} = H_v^\mathrm{T}+H_a^\mathrm{T},
    \end{split}
\end{equation}
where $H_a^\mathrm{V}, H_v^\mathrm{A} \subseteq H^\mathrm{VA}$ (moreover, $H^\mathrm{VA}=H_a^\mathrm{V} \cup H_v^\mathrm{A}$), $H_a^\mathrm{V}$ and $H_v^\mathrm{A}$ are the node feature matrices updated by improved GAT in $\mathrm{V}$-$\mathrm{A}$ graph; the rest can be deduced by analogy, $H_t^\mathrm{V}, H_v^\mathrm{T} \subseteq H^\mathrm{VT}$, and $H_t^\mathrm{A}, H_a^\mathrm{T} \subseteq H^\mathrm{AT}$; $H^\mathrm{V}$, $H^\mathrm{A}$, and $H^\mathrm{T}$ are the visual, acoustic, and textual feature matrices through multimodal fusion, respectively.

\subsection{Emotion prediction}
After multimodal fusion, we apply the concatenation operation to $H^\mathrm{V}$, $H^\mathrm{A}$, and $H^\mathrm{T}$. Finally, we obtain the final feature matrix $H$ carrying the three modal information. $H$ is computed by the following way:
\begin{equation}
    \label{eq:last_concat}
    \begin{split}
    H = \mathrm{W}_{vat}[H^\mathrm{V} \Vert H^\mathrm{A} \Vert H^\mathrm{T}],
    \end{split}
\end{equation}
where $\Vert$ is the concatenation operation and $\mathrm{W}_{vat}$ is the learnable parameter. We take $h_i \in H$ as the input to a fully connected network for emotion prediction:
\begin{equation}
    \label{eq:predict}
    \begin{split}
        &z_i = \mathrm{ReLU}(\mathrm{W}_{c}h_i+\mathrm{b}_{c}),\\
        &p_i = \mathrm{Softmax}(\mathrm{W}'_{c}z_i+\mathrm{b}'_{c}),\\
        &\hat{y}_i = \mathop{\mathrm{argmax}}_\mathrm{k}(p_i[\mathrm{k}]),
    \end{split}
\end{equation}
where $h_i \in H$ is the final feature vector of the $i$-th utterance $u_i$, and $\mathrm{ReLU}$ denotes the non-linear activation function; $p_i$ denotes the probability distribution of predicted emotion of $u_i$, and $\mathrm{Softmax}$ denotes the softmax function; $\hat{y}_i$ denotes the predicted emotion; $\mathrm{W}_{c}$, $\mathrm{W}'_{c}$, $\mathrm{b}_{c}$, and $\mathrm{b}'_{c}$ are the trainable parameters.

\subsection{Objective training}
We employ cross-entropy loss as objective function to train the proposed model, which can be formalized as:
\begin{equation}
    \label{eq:loss}
    \mathrm{Loss} = - \frac {1}{\sum_{t=0}^{N-1} n(t)} \sum_{i=0}^{N-1}\sum_{j=0}^{n(i)-1} y_{ij} \log p_{ij} + \lambda  \Vert \mathrm{W}_{ls} \Vert_2,
\end{equation}
where $N$ is the number of all conversations in the dataset, and $n(i)$ is the number of utterances of the $i$-th conversation; $y_{ij}$ denotes the ground-truth emotion of the $j$-th utterance in the $i$-th conversation, and $p_{ij}$ denotes the probability distribution of predicted emotion of the $j$-th utterance in the $i$-th conversation; $\lambda$ is the L2-regularizer weight, and $\mathrm{W}_{ls}$ is the set of all learnable parameters. 

\section{Experiments}\label{sec:experiments}
This section focuses on several experiment-related contents, including the baseline methods for comparison, the public benchmark datasets, and the detailed experimental settings.

\subsection{Baseline methods}
To demonstrate the effectiveness of GraphMFT in multimodal modeling, we compare it with some baseline models. The baseline methods we employ are as follows. \textbf{TFN}~\cite{zadeh2017tensor} allows dynamic modeling of intra- and inter-modality in an end-to-end manner. In this model, intra-modality is modeled by three modal embedding subnetworks, and inter-modality is modeled by a method called tensor fusion. \textbf{MFN}~\cite{zadeh2018memory} fuses multi-views information and unifies the multimodal features, but it ignores context-aware dependencies and speaker-aware dependencies. \textbf{BC-LSTM}~\cite{poria2017context} exploits textual, visual, and acoustic modalities to recognize emotion, and it captures multimodal information and extracts context-dependent features through a contextual LSTM network. \textbf{DialogueRNN}~\cite{majumder2019dialoguernn} tracks the individual party states in the conversation for emotion classification, where the characteristic of the speaker is taken into account for each incoming utterance to provide it with a more fine-grained context. \textbf{DialogueGCN}~\cite{ghosal2019dialoguegcn} is a graph-based model, where the conversation is regarded as a graph, which can efficiently capture long-range contextual information. We directly concatenate multimodal features to implement multimodal setting. \textbf{DialogueCRN}~\cite{hu2021dialoguecrn} investigates cognitive factors for ERC, where the cognitive phase is introduced to extract and fuse emotional clues from context and emotional clues are successfully utilized to better classify emotion states. The multimodal features are concatenated to achieve multimodal setting. \textbf{MMGCN}~\cite{hu2021mmgcn} designs a graph which can model the intra-speaker context and inter-modal dependencies to utilize multi-modal information as well as long-distance contextual information.

\subsection{Datasets}
\begin{table*}[htbp]
    \centering
    \renewcommand{\arraystretch}{1.2}
    \scriptsize
    \setlength{\tabcolsep}{4.3pt}
    \caption{The statistics of two public benchmark datasets adopted. The left shows the number of conversations, utterances, and classes. The right shows the distribution of sample size for each emotion.}
    \begin{tabular}{c|ccc|ccc|c||c|c|c|c|c|c|c}
    \hline
     &\multicolumn{3}{c|}{*Conversations} & \multicolumn{3}{c|}{*Utterances} & \multirow{2}[0]{*}{*Classes}  &\multirow{2}[0]{*}{*Neutral} &\multirow{2}[0]{*}{*Surprise} &\multirow{2}[0]{*}{*Fear} &\multirow{2}[0]{*}{*Sadness} &\multirow{2}[0]{*}{*Joy} & \multirow{2}[0]{*}{*Disgust}&\multirow{2}[0]{*}{*Angry}\\
     & \textit{train} & \textit{valid} & \textit{test} & \textit{train} & \textit{valid} & \textit{test} & &  & &  & &  &  &\\
    \hline
    MELD  &1039  & 114   & 280   & 9989  & 1109  & 2610  & 7& 6436  & 1636 & 358 & 1002 &2308 &361 &1607 \\
    IEMOCAP & \multicolumn{2}{c}{120} & 31    & \multicolumn{2}{c}{5810} & 1623  & 6 & 648  & 1084 & 1708 & 1103 &1041 &1849 & \\
    \hline
     & & & & & & & &*Happy &*Sad &*Neutral & *Angry &\makecell{*Excited} &\makecell{*Frustrated}& \\
    \hline
    \end{tabular}%
    \label{tab:datasets}%
\end{table*}
We conducted extensive experiments on two widely used public datasets, MELD~\cite{poria2018meld} and IEMOCAP~\cite{busso2008iemocap}. They are both multimodal benchmark datasets containing visual, acoustic and textual modalities. The statistics of the two datasets adopted are shown in Table~\ref{tab:datasets}.

\textbf{MELD} dataset is a multi-party and multimodal dataset for ERC task. It includes textual, visual and acoustic modalities information and over 1400 conversations, a total of 13700 utterances, which are categorized to seven emotions: \textit{Neutral}, \textit{Surprise}, \textit{Fear}, \textit{Sadness}, \textit{Joy}, \textit{Disgust}, and \textit{Angry}. It has three or more speakers in each conversation and is collected from the TV show \textit{Friends}.

\textbf{IEMOCAP} dataset is an ERC dataset containing multimodal audio-video-text conversations from ten professional actors. It contains 151 conversations, a total of 7433 dyadic utterances, where each utterance corresponds to an emotion label. The emotion labels of these utterances are divided into six categories, including \textit{Happy}, \textit{Sad}, \textit{Neutral}, \textit{Angry}, \textit{Excited}, and \textit{Frustrated}.

It should be noted that the MELD dataset differs from the IEMOCAP dataset in that MELD suffers from a severe class-imbalanced problem. The right of Table~\ref{tab:datasets} shows the distribution of sample size for each class in both datasets. It can be obviously seen that the sample size of \textit{Fear} and \textit{Disgust} in MELD is much smaller than that of other emotions, and such class-imbalanced problem poses a great challenge to emotion classification.

\subsection{Detailed settings}
We perform all experiments on NVIDIA GeForce RTX 3090, and the operating system is Ubuntu 20.04. Pytorch is used as deep learning framework and its version is 1.12.0. All experiments are implemented with Python 3.9.12. We choose AdamW as the optimizer to train proposed model. The maximum epoch is set to 100, the L2 regularization factor and dropout rate are set to 0.00001 and 0.5, respectively. For MELD dataset, the number of network layers is 3; the learning rate is 0.00002; the batch size is 32. For IEMOCAP dataset, the number of network layers is 5; the learning rate is 0.00001; the batch size is 16. Table~\ref{tab:implementation} describes the partial hyperparameter settings. To evaluate the superiority of our proposed GraphMFT, we apply weighted-average F1 score and accuracy score as evaluation metrics.
\begin{table}[htbp]
    \centering
    \renewcommand{\arraystretch}{1.2}
    \setlength{\tabcolsep}{1.5pt}
    \caption{Description of the partial hyperparameter settings.}
    \begin{tabular}{c|c|c|c|c|c}
    \hline
    Dataset & \makecell{Number of\\layers} & \makecell{Learning\\rate} & \makecell{Batch\\size} & \makecell{Regulariza-\\tion factor} & \makecell{Dropout\\rate}\\
    \hline
    MELD  & 3  & 2e-05 & 32 & 1e-05 &0.5\\
    IEMOCAP &  5  & 1e-05 & 16 & 1e-05 &0.5\\
    \hline
    \end{tabular}%
    \label{tab:implementation}%
\end{table}

\section{Results and analysis}\label{sec:results}
We report experimental results to evaluate the proposed GraphMFT in this section. First, GraphMFT is compared with all baseline methods as a whole. Then, the influence of different settings on our GraphMFT is discussed. Finally, we also carry out limitations analysis of our approach.

\subsection{Overall results}
The experimental results of all models are recorded in Table~\ref{tab:overall}. It can be significantly seen that our GraphMFT outperforms all baseline models. On the IEMOCAP dataset, the accuracy score and weighted-average F1 score of GraphMFT are 67.90\% and 68.07\%, respectively, which are distinctly higher than those of SOTA method (i.e., MMGCN). DialogueCRN is another strong baseline model in addition to MMGCN. From Table~\ref{tab:overall}, we can conclude that F1 score of GraphMFT on the IEMOCAP dataset is 2.73\% higher than that of DialogueCRN. We record F1 score for each emotion on the IEMOCAP dataset. We found that, except for \textit{Happy} and \textit{Frustrated}, GraphMFT's F1 scores for other emotions have significant improvement compared to other models. Notably, GraphMFT performs better for \textit{Sad} and \textit{Excited} than for other emotions, achieving F1 scores of 83.12\% and 76.92\%, respectively. In addition, the accuracy score of GraphMFT on the MELD dataset is improved by 1.99\% relative to that of MMGCN. In general, the proposed GraphMFT shows superior performance than MMGCN and DialogueCRN in multimodal fusion.
\begin{table*}[htbp]
    \centering
    \renewcommand{\arraystretch}{1.2}
    \setlength{\tabcolsep}{8pt}
    \caption{The overall results of all models. Evaluation metrics contain $\mathrm{Acc}$, $\mathrm{F1}$, and $\mathrm{wa}$-$\mathrm{F1}$, which denote accuracy score (\%), F1 score (\%), and weighted-average F1 score (\%), respectively. Bold font indicates the best performance.}
    \begin{tabular}{c|cccccc|cc||cc}
    \hline
    \multicolumn{1}{c|}{\multirow{3}{*}{Model}} & \multicolumn{8}{c||}{IEMOCAP} & \multicolumn{2}{c}{MELD} \\\cline{2-11}          
    & \textit{Happy} & \textit{Sad}   & \textit{Neutral} & \textit{Angry} & \textit{Excited} & \textit{Frustrated} & \multirow{2}{*}{Acc} & \multirow{2}{*}{wa-F1} & \multirow{2}{*}{Acc} & \multirow{2}{*}{wa-F1} \\ \cline{2-7}
    & F1 & F1   & F1 & F1 & F1 & F1 &  &  &  &  \\
    \hline
      TFN & 38.29 & 64.12 & 51.43 & 55.34 & 57.95 & 57.58 & 56.12 & 55.54 & 59.77 & 57.01 \\
      BC-LSTM & 32.63 & 70.34 & 51.14 & 63.44 & 67.91 & 61.06 & 59.58 & 59.10  & 59.62 & 56.80 \\
      MFN & 47.19 & 72.49 & 55.38 & 63.04 & 64.52 & 61.91 & 60.14 & 60.32  & 59.93 & 57.29 \\
      DialogueRNN & 33.18 & 78.80  & 59.21 & 65.28 & 71.86 & 58.91 & 63.40  & 62.75 & 60.31 & 57.66 \\
      DialogueGCN & 47.10  & 80.88 & 58.71 & 66.08 & 70.97 & 61.21 & 65.54 & 65.04  & 58.62 & 56.36 \\
      DialogueCRN & \textbf{51.59} & 74.54 & 62.38 & 67.25 & 73.96 & 59.97 & 65.31 & 65.34  & 59.66 & 56.76 \\
      MMGCN & 45.45 & 77.53 & 61.99 & 66.67 & 72.04 & \textbf{64.12} & 65.56 & 65.71 & 59.31 & 57.82 \\
      \hline
      GraphMFT  & 45.99 & \textbf{83.12} & \textbf{63.08} & \textbf{70.30}  & \textbf{76.92} & 63.84 & \textbf{67.90} & \textbf{68.07}  & \textbf{61.30} & \textbf{58.37} \\
      \hline
    \end{tabular}%
    \label{tab:overall}%
\end{table*}%

\subsection{Comparison of vanilla and improved GATs}
To demonstrate the validity of the improved GAT, we replace the improved GAT module in GraphMFT with the vanilla GAT. We compare the experimental results with vanilla GAT and improved GAT on these two datasets, as displayed in Table~\ref{tab:two_gat}. It is observed that our GraphMFT with improved GAT module achieves a 1.60\% improvement of weighted-average F1 score on the IEMOCAP dataset in comparison to the model with vanilla GAT. The similar results appear on the MELD dataset. For instance, the model using vanilla GAT module achieves an accuracy score of 60.19\% on the MELD dataset, which is 1.11\% lower than GraphMFT using improved GAT. Therefore, our improved GAT can be an effective module to promote the performance of emotion classification.
\begin{table}[htbp]
    \centering
    \renewcommand{\arraystretch}{1.2}
    \setlength{\tabcolsep}{8pt}
    \caption{The experimental results with vanilla GAT and improved GAT.}
      \begin{tabular}{c|cc||cc}
      \hline
      \multicolumn{1}{c|}{\multirow{2}[1]{*}{Module}} & \multicolumn{2}{c||}{IEMOCAP}  & \multicolumn{2}{c}{MELD} \\
      \cline{2-5}
            & Acc & wa-F1 & Acc & wa-F1 \\
      \hline
      Vanilla GAT     & 66.97  & 66.47  & 60.19  & 57.90  \\
      Improved GAT     & \textbf{67.90}  & \textbf{68.07}  & \textbf{61.30}  & \textbf{58.37}  \\
      \hline
      \end{tabular}%
    \label{tab:two_gat}%
\end{table}%

\subsection{Impact of number of network layers}
\begin{figure}[htbp]
    \centering
    \includegraphics[width=3.0in]{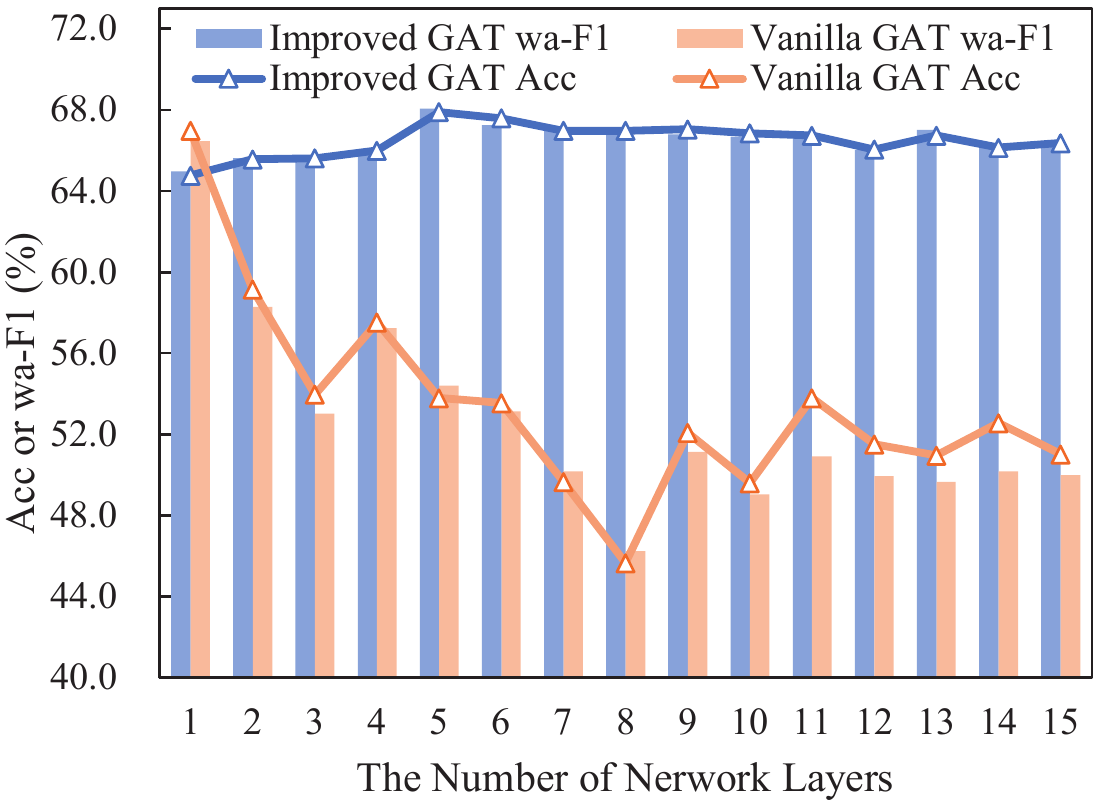}
    \caption{Experimental results for testing different number of network layers on the IEMOCAP dataset.}
    \label{fig:layer}
\end{figure}

When the number of network layers reaches to a specific threshold, the performance of GNNs drops dramatically if the number of network layers is increased further. This is notorious over-smoothing problem in graph models~\cite{kipf2016semi}. We make an improvement to standard GAT in Section~\ref{sec:multi-graph_interactions} based on the idea of ResGCN. In order to exhibit that the proposed approach alleviates the over-smoothing problem, we replace the improved GAT module in GraphMFT with the vanilla GAT, and provide a comparison of the experimental results using the vanilla GAT and the improved GAT. Figure~\ref{fig:layer} shows the effect of the number of network layers on the model performance. Note that the reported experimental results are based on the ICMOCAP dataset. 

We can visualize that the performance of the model replaced with the vanilla GAT module decreases sharply as the number of network layers increases. When a certain threshold is reached, the performance of the model fluctuates around a lower value as the number of network layers is varied. Such result prohibits the possibility to improve the expressiveness of the model through increasing the number of network layers. As we expected, the performance of GraphMFT utilizing the improved GAT increases with the number of network layers in the beginning stage; when the number of network layers reaches a certain threshold, the performance of GraphMFT slowly decreases as the number of layers increases. Distinct from the vanilla GAT module, GraphMFT with improved GAT has a very weak performance degradation.

\subsection{Impact of speaker embedding}
Previous baseline approaches, such as MMGCN, proof that the characteristics of speakers play an essential part in ERC tasks. Table~\ref{tab:speaker} shows the effect of adding or not adding speaker embeddings on the performance of GraphMFT. On the IEMOCAP dataset, the weighted-average F1 score without speaker embedding is 65.61\%, which is 2.46\% lower than the score with speaker embedding; the accuracy score with speaker embedding is increased by 2.16\% in comparison with that without speaker embedding. On the MELD dataset, the performance of adding speaker embedding is also higher than that of not adding speaker embedding. Therefore, we can conclude that adding speaker features can improve the performance of ERC task. In addition, we realize that GraphMFT without speaker information and MMGCN with speaker information have comparable performance, which indicates that the proposed GraphMFT has excellent multimodal modeling capability.
\begin{table}[htbp]
    \centering
    \renewcommand{\arraystretch}{1.2}
    \setlength{\tabcolsep}{6pt}
    \caption{The effect of adding or not adding speaker embeddings on the performance of GraphMFT. w/o (w) indicates without (with) speaker embeddings.}
      \begin{tabular}{c|cc||cc}
      \hline
      \multicolumn{1}{c|}{\multirow{2}[1]{*}{Speaker Embedding}} & \multicolumn{2}{c||}{IEMOCAP}  & \multicolumn{2}{c}{MELD} \\
      \cline{2-5}
            & Acc & wa-F1 & Acc & wa-F1 \\
      \hline
      w/o     & 65.74  & 65.61  & 61.03  & 58.04  \\
      w     & \textbf{67.90}  & \textbf{68.07}  & \textbf{61.30}  & \textbf{58.37}  \\
      \hline
      \end{tabular}%
    \label{tab:speaker}%
\end{table}%

\subsection{Comparison of different modal settings}
To investigate the effect of different modal settings on the performance of ERC model, we report in Table~\ref{tab:modal_set} the experimental results for the model with these settings. We can clearly see that the performance of three-modality-based models outperforms that of all two-modality-based models, indicating that more modalities contribute to the accuracy of emotion recognition. For the two-modal settings, the performance of the model based on visual and acoustic modalities is the worst, while the model with a combination of acoustic and textual modalities performs the best. Therefore, it can be concluded that the acoustic and textual modalities have more contributions to GraphMFT compared to the visual modality.
\begin{table}[htbp]
    \centering
    \renewcommand{\arraystretch}{1.2}
    \setlength{\tabcolsep}{8pt}
    \caption{Experimental results for different modal settings. V, A, and T indicate visual, acoustic, textual modalities, respectively.}
      \begin{tabular}{c|cc||cc}
      \hline
      \multicolumn{1}{c|}{\multirow{2}[1]{*}{Modal Settings}} & \multicolumn{2}{c||}{IEMOCAP}  & \multicolumn{2}{c}{MELD} \\
      \cline{2-5}
            & Acc & wa-F1 & Acc & wa-F1 \\
      \hline
      V-A     & 52.87  & 51.95  & 48.70  & 43.37  \\
      V-T     & 61.49  & 61.66  & 60.57  & 57.63  \\
      A-T     & 66.79  & 66.65  & 60.84  & 57.51  \\
      V-A-T   & \textbf{67.90}  & \textbf{68.07}  & \textbf{61.30}  & \textbf{58.37}  \\
      \hline
      \end{tabular}%
    \label{tab:modal_set}%
\end{table}%

\subsection{Impact of number of contextual nodes}
\begin{figure*}[htbp]
    \centering
    \subfloat[Results for different $\mathcal{P}$ ($\mathcal{F}$) on the IEMOCAP dataset]{\includegraphics[width=2.5in]{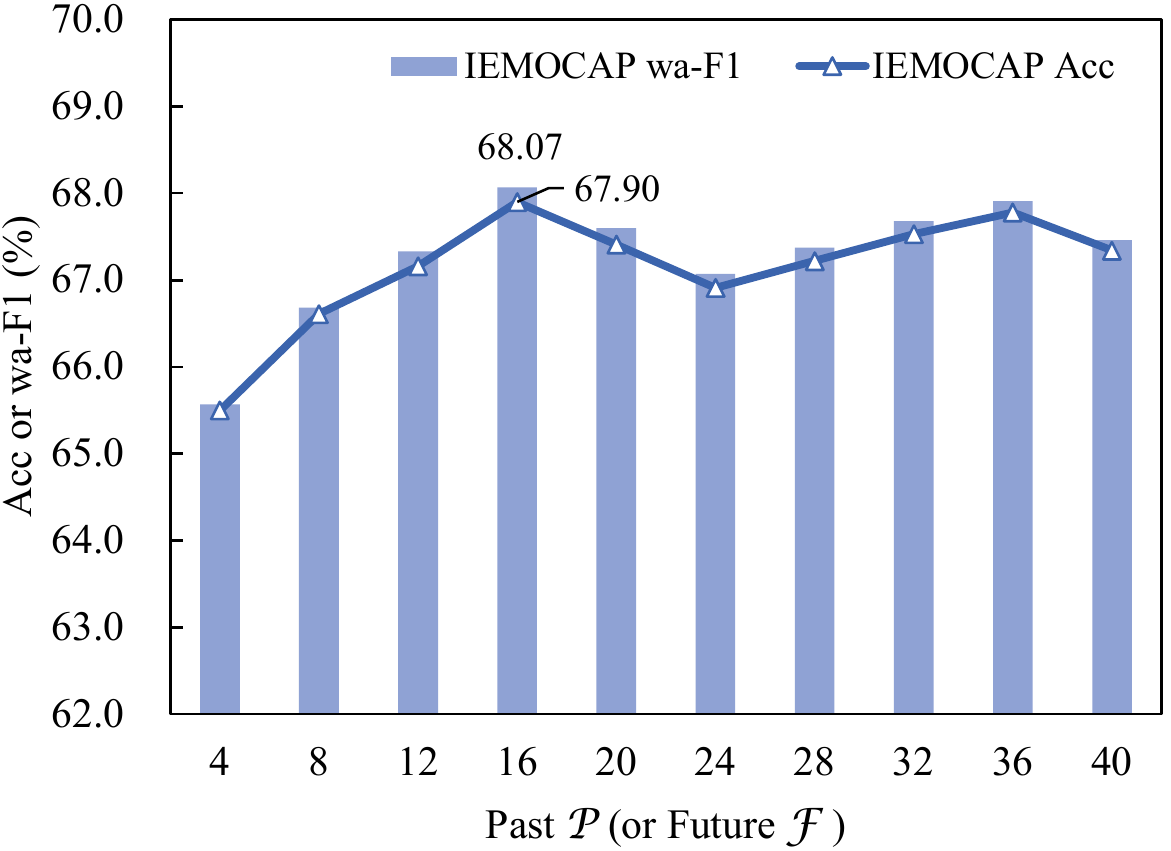}%
    \label{fig:past0}}
    \hfil
    \subfloat[Results for different $\mathcal{P}$ ($\mathcal{F}$) on the MELD dataset]{\includegraphics[width=2.5in]{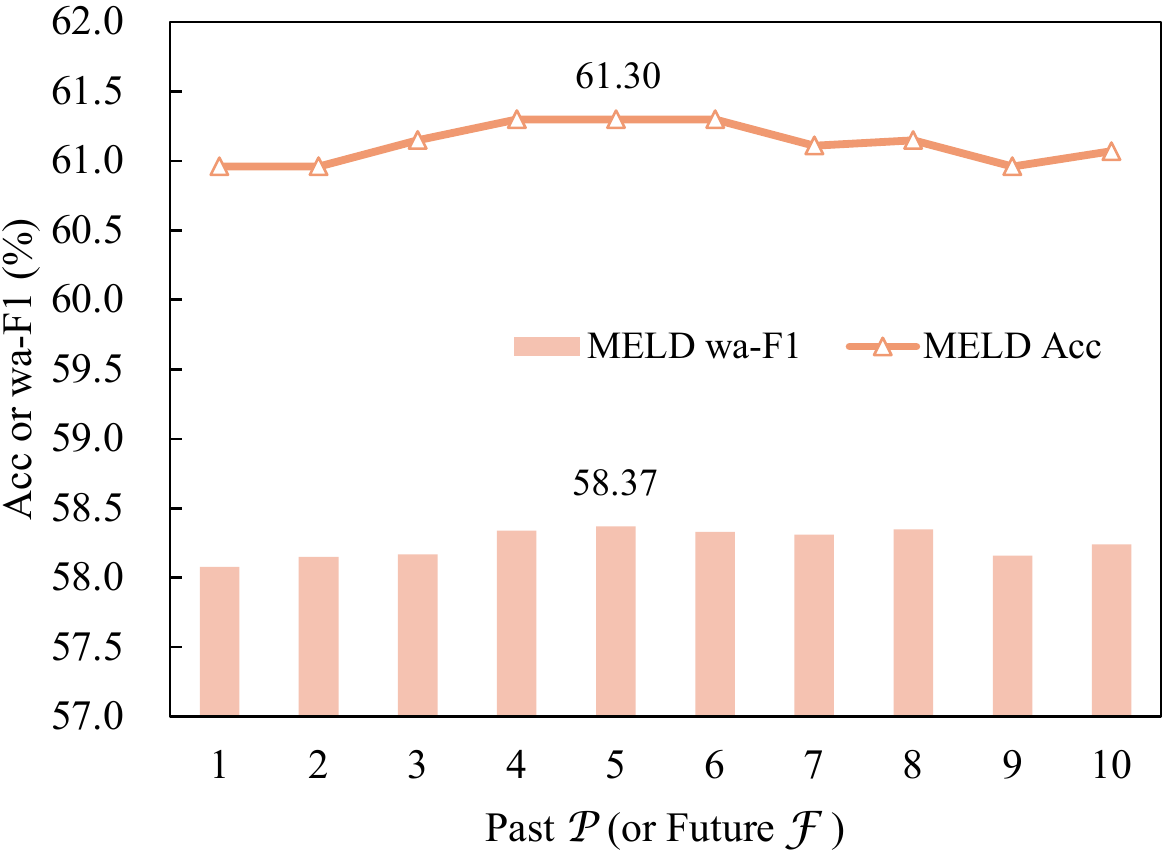}%
    \label{fig:past1}}
    \caption{The effect of different combinations $(\mathcal{P}, \mathcal{F})$ on the performance of our model. Here, $\mathcal{P}$ (or $\mathcal{F}$) denotes the number of past (or future) contextual nodes of current node.
    }
    \label{fig:past}
\end{figure*}

When we create intra-modal edges, current node needs to connect the past $\mathcal{P}$ and the future $\mathcal{F}$ contextual nodes. This section investigates the impact of various $(\mathcal{P}, \mathcal{F})$ combinations on the performance of our model. Specifically, we set the following $(\mathcal{P}, \mathcal{F})$ combinations: (a) for the IEMOCAP dataset, the combinations are set to $(4,4)$, $(8,8)$, $\cdots$, $(40,40)$; (b) for the MELD dataset, the combinations are set to $(1,1)$, $(2,2)$, $\cdots$, $(10,10)$. Figure~\ref{fig:past} depicts the effect of different $(\mathcal{P}, \mathcal{F})$ combinations on accuracy scores and weighted-average F1 scores of GraphMFT. It can be concluded from Figure~\ref{fig:past0} that the performance of GraphMFT on the IEMOCAP dataset increases with increasing $\mathcal{P}$ or $\mathcal{F}$; when reaching a certain threshold $(16,16)$, both accuracy score and F1 score of GraphMFT start to fluctuate up and down. Figure~\ref{fig:past1} shows the performance variation of proposed model on the MELD dataset. What can be expected is that the variation pattern of the performance on the MELD dataset and the IEMOCAP dataset are nearly identical. Another interesting phenomenon is that GraphMFT requires $16$ contextual nodes for optimal performance on the IEMOCAP dataset, while only $5$ contextual nodes are required on the MELD dataset. The probable reason is that some neighboring utterances in the MELD dataset are not contiguous in the actual scenario, and thus the long-range context modeling is not required.

\subsection{Limitations}
\begin{figure*}[htbp]
    \centering
    \subfloat[Confusion matrix of proposed GraphMFT]{\includegraphics[width=2.5in]{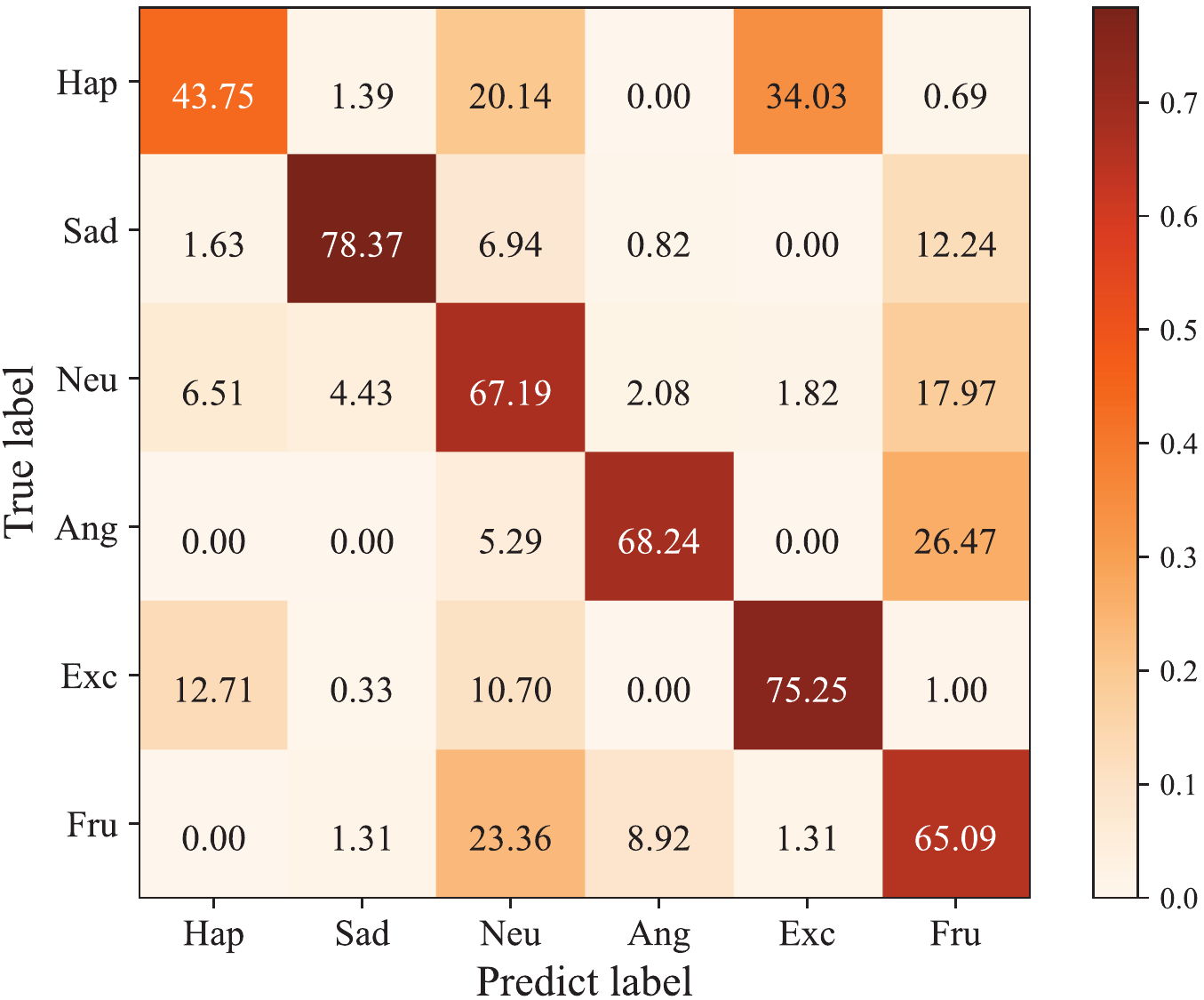}%
    \label{fig:error0}}
    \hfil
    \subfloat[Confusion matrix of previous MMGCN]{\includegraphics[width=2.5in]{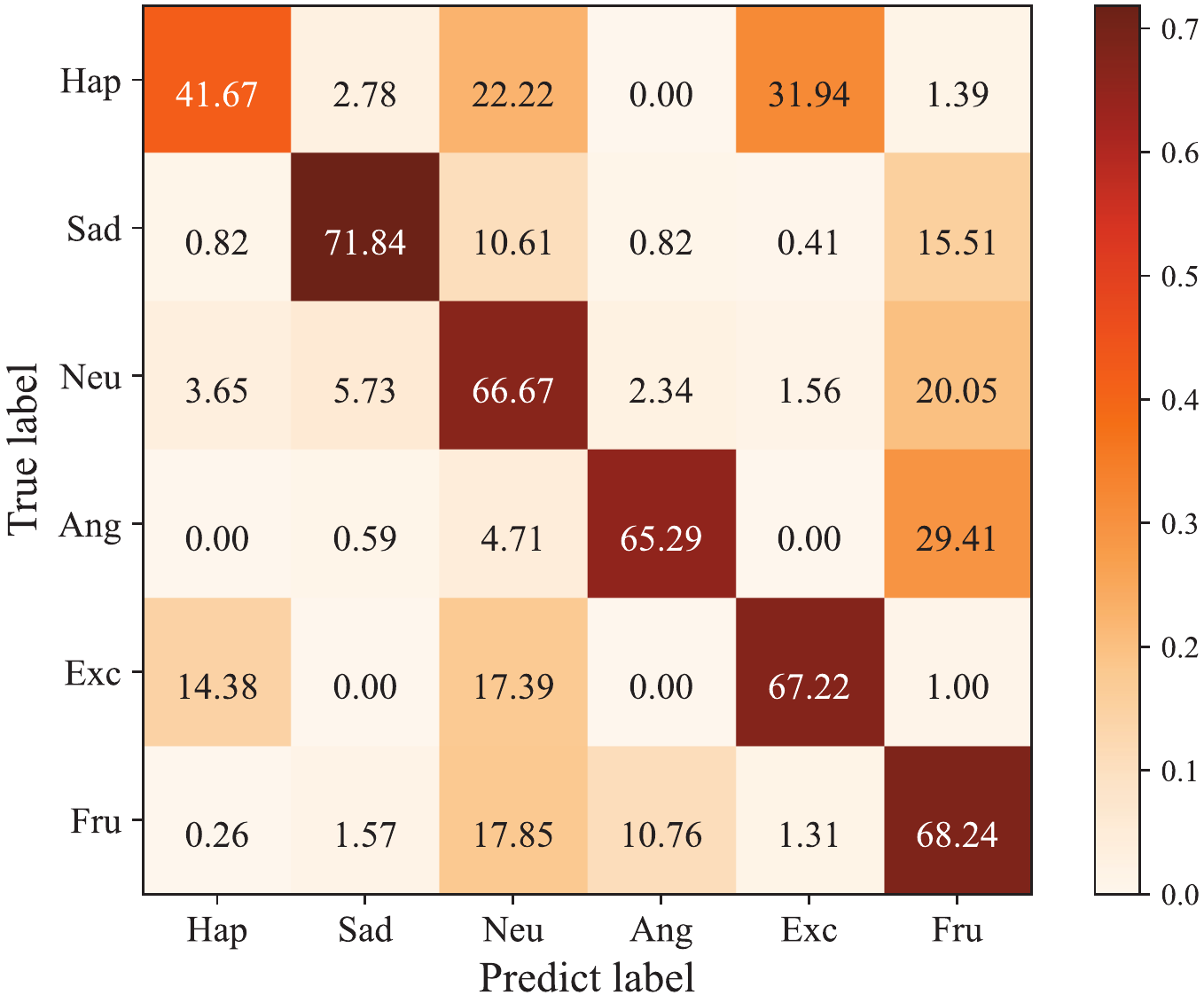}%
    \label{fig:error1}}
    \caption{The confusion matrices of proposed GraphMFT and previous MMGCN on the IEMOCAP dataset. GraphMFT, like MMGCN, faces the problems of \textit{Similar-Emotion} and \textit{non-Neutral emotion}.
    }
    \label{fig:error}
\end{figure*}
As shown in Figure~\ref{fig:error}, we discuss the limitations of current ERC models by leveraging two confusion matrices. One of the challenges of ERC task is the \textit{Similar Emotion} problem, meaning that many models have difficulty distinguishing similar emotions. Like existing models, our GraphMFT sometimes misclassifies similar emotions. As shown in Figure~\ref{fig:error0}, GraphMFT incorrectly detects ground-truth emotion \textit{Angry} as \textit{Frustrated} with a 26.47\% probability; the ground-truth emotion \textit{Excited} is incorrectly detected as \textit{Happy} with a 12.71\% probability. Another annoying problem in ERC task is that models are eager to recognize many emotions as \textit{Neutral}, which we refer to as the \textit{non-Neutral Emotion} problem. Most existing ERC models rely on text modality. Thus, when some non-\textit{Neutral} utterances such as "Okay." and "Yeah." appear in the conversation, the model may detect them as \textit{Neutral} directly. As shown in Figure~\ref{fig:error0}, the ground-truth emotion \textit{Happy} is identified as \textit{Neutral} with a probability of 20.14\%.

Fortunately, the proposed GraphMFT can alleviate \textit{Similar Emotion} problem compared to the baseline models such as MMGCN. By comparing Figure~\ref{fig:error0} and Figure~\ref{fig:error1}, it can be concluded that GraphMFT reduces the probability that \textit{Angry} is identified as \textit{Frustrated} from 29.41\% of MMGCN to 26.47\%; similarly, the probability that \textit{Excited} is identified as \textit{Happy} is reduced from 14.38\% of MMGCN to 12.71\% of GraphMFT. In addition, GraphMFT also mitigates \textit{non-Neutral Emotion} problem. For example, the probability that \textit{Sad} is classified as \textit{Neutral} is reduced by 3.67\% from 10.61\% of MMGCN; GraphMFT reduces the probability that \textit{Excited} is classified as \textit{Neutral} from 17.39\% to 10.70\% of MMGCN. Hence, GraphMFT can compensate textual modal misinformation by visual and acoustic modal information, and adequately capture multimodal complementary information to improve the ability of emotion recognition.

\section{Summary}\label{sec:summary}
To overcome the challenges of existing graph-based multimodal ERC models, we propose a novel multimodal fusion method, namely Graph network based Multimodal Fusion Technique (GraphMFT). The proposed GraphMFT models each conversation as three heterogeneous graphs, i.e., V-A graph, V-T graph, and A-T graph. Each graph includes data from only two modalities to diminish the difficulty of multimodal fusion. In addition, to handle the over-smoothing problem of graph neural networks, we improve the vanilla graph attention network inspired by ResGCN. We evaluate our proposed GraphMFT on two benchmark datasets that are widely used by most ERC models. Experimental results demonstrate that our proposed model can effectively perform intra- and inter-modal interactions to extract contextual information and complementary information. By comparing with the previous baseline model, GraphMFT prevails with a significant superiority.

Future studies could fruitfully and continuously explore the application of multimodal machine learning for conversational emotion recognition. Meanwhile, with the relational modeling capability of graph neural networks, we hope to further extend the graph-based models to the field of multimodal dialogue generation.

\section*{CRediT authorship contribution statement}
\textbf{Jiang Li:} Conceptualization, Methodology, Data Curation, Software, Validation, Formal analysis, Investigation, Visualization, Writing - original draft, Project Administration, Writing - review \& editing. \textbf{Xiaoping Wang:} Supervision, Writing - review \& editing, Funding Acquisition. \textbf{Guoqing Lv:} Investigation, Writing - review \& editing. \textbf{Zhigang Zeng:} Supervision, Funding Acquisition.

\section*{Data availability}
Data will be made available on request.

\section*{Declaration of competing interest}
The authors declare that they have no known competing financial interests or personal relationships that could have appeared to influence the work reported in this paper.

\section*{Acknowledgments}
This work was supported in part by the National Natural Science Foundation of China under Grant 62236005, 61936004, and U1913602.

\bibliographystyle{elsarticle-num}
\balance
\bibliography{graphmft}

\end{document}